\documentclass[12pt]{article}
\usepackage{epsf}
\title{Heavy quark distribution
function in hadrons} \vspace{2cm}
\author{A.G.Oganesian\\
\\
Institute of Theoretical and Experimental Physics, Moscow, Russia}
\date{}

\begin{document}
\maketitle

\date{}

\maketitle

\newcommand{\be}{\begin{equation}}
\newcommand{\ee}{\end{equation}}

\def\la{\mathrel{\mathpalette\fun <}}
\def\ga{\mathrel{\mathpalette\fun >}}

\def\fun#1#2{\lower3.6pt\vbox{\baselineskip0pt\lineskip.9pt

\ialign{$\mathsurround=0pt#1\hfil##\hfil$\crcr#2\crcr\sim\crcr}}}

\maketitle

\begin{abstract}
\noindent The moments of the heavy quark-parton distribution
functions in a heavy pseudoscalar meson, obtained in QCD sum
rules, are expanded in the inverse heavy quark. Comparison with
the finite mass results reveals that while the heavy mass
expansion works reasonably well for the $b$ quark, one has to take
into account terms of higher than $(1/m_c)^2$ order for the $c$
quark.
\end{abstract}



\markboth{\large \sl \underline{A.~G.~OGANESIAN} \hspace*{2cm}
HSQCD 2008} {\large \sl \hspace*{1cm} TEMPLATE FOR THE HSQCD 2008
PROCEEDINGS}

In this talk I will very shortly report the result, obtained
recently in my paper \cite{ao}. The problem is that  there exist a
large number of different theoretical models and approaches
usually used to describe the heavy quark fragmentation functions
(FF)(see, for example $[2-6]$), and expansions, based on the
heavy-quark mass limit or on HQET are widely used in many of them
(e.g. \cite{a4}). So the question arise, does the expansion work
good, especially for charm case. In \cite{ao} it was offered to
check the applicability (and accuracy) of this expansion using the
moments of the heavy-quark structure obtained long ago from QCD
sum rules in \cite{n1}. In this approach the moments of the
heavy-quark parton distribution functions $ M_n^c=\int\limits_0^1
dx x^{n-1} c(x,Q^2) $ were directly calculated in QCD sum rules
and  FF were estimated, following the familiar relation \cite {gl}
( see also \cite{kart} and \cite{n1} for details).  The advantage
of this method is the possibility  to estimate the heavy parton
distribution functions (hence, also fragmentation functions) in
full QCD, in terms of universal parameters, such as quark masses
and condensate densities. One can use this result to fix the
parameters of different models, as was discussed in \cite{n1}.

There is no time to discuss the method itself, one can look it in
the paper  \cite{n1} and also \cite{ao}. So I very briefly remind
the main points of the method. One starts from the four-point
correlator:

\be \Pi_{\mu\nu}  =i\int e^{ip_1x+iq_1y-ip_2z}d^4xd^4yd^4z\langle
0\mid T \left \{j_5(x)
j^{\mbox{em}}_{\mu}(y)j^{\mbox{em}}_{\nu}(0)j_5^\dagger(z)\right
\}\mid 0 \rangle, \ee where
$j^{\mbox{em}}_{\mu}=\bar{c}\Gamma_{\mu}c $, and heavy-light
pseudoscalar currents $j_5=\bar{c}\Gamma_5u $ interpolate $D$
meson. The above correlator is considered in the deep spacelike
region at
 $t=(p_1-p_2)^2=0$, $p_i^2,q_i^2<0$. We take into account the
contributions of the unit operator (bare loop), quark condensate
$a=- (2\pi)^4\langle 0\mid\bar{\psi}\psi\mid 0\rangle$ and
quark$-$gluon condensate $\langle 0\mid
\bar{\psi}G_{\mu\nu}\sigma^{\mu\nu}\psi\mid 0\rangle$. The gluon
and four-quark condensate contributions are neglected, because the
estimates show, that they are very small. We express this
amplitude in terms of double dispersion relation over $p_1^2$ and
$p_1^2$, then, us usual,  saturate the amplitude by hadronic
states and equate it to the result of OPE and  perform two
independent Borel transformations on $p_1^2$ and $p_2^2$. (We will
equate the Borel masses $M_{1B}^2 = M_{2B}^2 =M_B^2$  only in the
final sum rule). Following \cite{n1}, we choose the value of the
variable $s=(p_1+q_1)^2$ in the unphysical region as
$s=m_D^2-Q^2$, and transform the integrals into  the moments of
the parton distribution. Finally the result, obtained in, is
\cite{n1}:

\be M_n=g_D^{-2}m_D^{-4}\Biggl ( \int\limits^{s_0}_{m^2}du_1
\int\limits^{z_+}_{z_-}dz
\frac{f_0e^{{-2(u_1-m_D^2)}/{M_B^2}}}{(1+z-m_D^2/Q^2)^{n+1}\pi^2}
+\frac{maLe^{{-2(m^2-m_D^2)}/{M_B^2}}}{4\pi^2\eta^{n+1}} +
R_6\Biggr)
 \ee \label{8}

Here, \be f_0=\frac{3}{8}(s_1-m^2)\Biggl[ \frac{
1+z+(s_1+2m^2)/Q^2}{((1+z-s_1/Q^2)^2+4s_1/Q^2 )^{-1/2}}
 -\frac{4m^2}{Q^2}\Biggl (1+\frac{4m^2}{Q^2}\Biggr
)^{-1/2}\Biggr] \label{9} \ee and
$$
R_6 = \frac{m m_0^2 a}{4\pi^2}\frac{1}{2M_B^2}\Biggl
[4/M_B^2-2m^2/M_B^4 +1/Q^2+
\frac{(n+1)}{\eta}\Biggl(\frac{2}{3Q^2} -1/M_B^2 -
\frac{2m^2}{M_B^2Q^2}\Biggr) -$$ \be
 -\frac{(n+1)(n+2)m^2}{Q^4\eta^2}
\Biggr ] \eta^{-n-1}e^{-2(m^2-m_D^2)/{M_B^2}}, \label{10} \ee

Here $\eta=(1-(m^2-m_D^2)/Q^2)$. The factor $L$ accounts for the
anomalous dimension of the quark condensate, normalized at point
$\mu$. We adopt the values of condensates \cite{giz,iof},
 normalizing them  at $\mu=1~ \mbox{GeV}$. In this case
the factor $L$ is close to the unity.

In \cite{n1},  it was shown that the first few moments of the sum
rules are well behaved and the following numerical results were
obtained at $Q_0^2=20~ \mbox{GeV}^2 $ ($Q_0^2=10~ \mbox{GeV}^2 $):
$M_2=0.85 (0.9), M_3=0.75(0.83), M_4=0.67(0.78)$ These moments can
be used as an input in the evolution equations, at some initial
point $Q_0^2 = (3-5)m_D^2$, hence it is possible to predict the
structure functions also at large $Q^2$.

Now we can answer to our question if it is possible to use the
heavy$-$ quark limit for the heavy quark parton distribution (and,
correspondingly, fragmentation function). For that, we expand the
sum rules  for the moments in the inverse mass of  the heavy
quark.

Substituting in Eq.(4) instead of the $g_D^2 m_D^4$  the sum rule
obtained from the two-point correlator  \cite{aliev}
 we can rewrite the sum rules for  $M_n$ as \be M_n=
\frac{R_0(n)+R_4(n)+R_6(n)}{K_0+K_4+K_6},
 \label{12} \ee where the
terms $R_d(n)$ in the numerator and $K_d$  in the denominator
originate from the dimension $d=0,4,6$ contributions to the OPE of
the four-point and two-point correlator, respectively. Following
\cite{n1} we adopt the relation $M^2_B/2=M^2$ between the Borel
masses in the four-point and two-point sum rules. Note that the
exponential factor containing $m_D^2$ cancels in the ratio of the
two sum rules.

To investigate the heavy$-$quark mass limit of (\ref{12}) we
employ the standard scaling relations for the heavy hadron mass,
continuum threshold, and Borel parameter: $ m_D=m+ \bar{\Lambda},$
 $s_0=m^2+2m\omega,$ $ M^2=2\tau m \label{13}$ and in addition
assume the scaling $Q^2=\gamma m_D^2$, where in the heavy$-$quark
limit  the parameters $\bar{\Lambda}$, $\omega$, $\tau$ and
$\gamma$ do not depend of the heavy quark mass. (One should note,
that $\bar{\Lambda}$ here has nothing common with well-known QCD
parameter $\Lambda_{QCD})$. In what follows we will use notation
$\delta= \bar{\Lambda}/m,$ $y=\bar{\Lambda}/\tau.$.

Expanding in powers of $1/m$ , we obtain the first three terms for
$R_0$:
$$
R_0(n) \simeq D_2 +(n+1)\delta\frac{2D_2-(2+\gamma)D_3}{\gamma}
+\delta^2(n+1)(n+2) \Biggl  [
\frac{2(n+2)+3\gamma}{\gamma^2(n+2)}D_2 -$$
 \be -
\frac{2}{\gamma}\Biggl (1+\frac{2(n+1)}{(n+2)\gamma}\Biggr )D_3 +
\frac{2D_4}{3\gamma^2}\Biggl(\gamma^2+4\gamma+3
+\frac{2\gamma}{(n+1)(n+2)}\Biggr ) \Biggr ]. \ee

Here, $D_n$ is defined as $D_n=E_n(\omega/\tau)/y^{n-2}$ and
$E_n(z)=\int\limits^{z}_{0}dze^{-z}z^{n}$ For the contribution of
$d=4 $ operators we obtain:
 \be R_4(n)
\simeq \frac{a}{12\tau^3}L\Biggl ( 1+2\delta(n+1)/\gamma
+\delta^2\frac{(n+1)(n+2)}{\gamma}[2/\gamma-3/(n+2)]\Biggr ). \ee

Finally, the contribution of $d=6 $ operators transforms into
(first three terms)
$$
 R_6(n)\simeq\frac{a}{12\tau^3}\frac{m_0^2}{16\tau^2} \Biggl [
-1+\frac{2\delta}{y}\Biggl(4-(n+1)[1+(y+2)/\gamma] \Biggr)-$$
 \be
-\frac{2\delta^2(n+1)}{y\gamma}
\Biggl((n+2)[2+(y+4)/\gamma+2/(y\gamma)]
-3y-24-(8/3)(2n+5)/y\Biggr)\Biggr]. \ee

And, finally, for the denominator$K=K_0+K_4+K_6$ in (\ref{12}) we
obtain after expansion \be K \simeq D_2 - 2D_3\delta -
4D_4\delta^2+ \frac{a}{12\tau^3}\Biggl( L
-\frac{m_0^2}{16\tau^2}\Biggr)
+\frac{a}{12\tau^3}\frac{m_0^2}{16\tau^2}\frac{4\delta}{y}. \ee

\begin{figure}
\epsfxsize=10cm \epsfbox{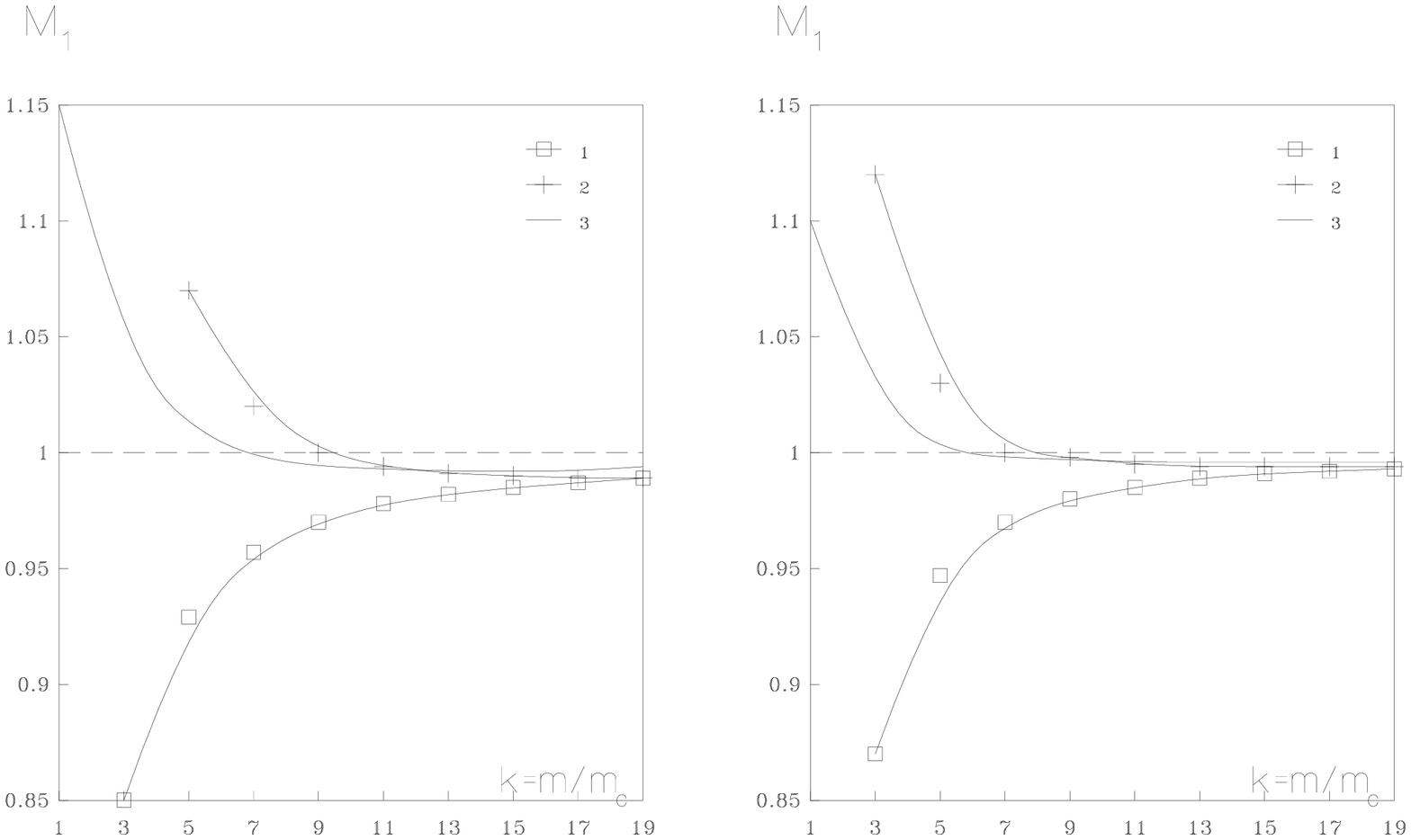} \caption{ The first moment as
a function of the quark mass $m$ in the units $k=m/m_c$. Curves
1,2,3 correspond to the first-and second-order expansion in the
inverse quark mass and to the exact answer, respectively. The
value of $\gamma=Q^2/M_D^2=2(4)$ is chosen in the left (right)
panel.}
\end{figure}

From these equations it is easy to calculate the actual value of
moments in the first and second orders of expansion and compare it
with the exact value to examine the accuracy of the expansion in
the inverse heavy quark mass. For numerical analysis we choose
$\bar{\Lambda}=0.6~ \mbox{GeV}$, $\omega=1.4~ \mbox{GeV},$ and
$\tau=0.6~ \mbox{GeV}$, as usual. The numerical results for
heavy$-$quark mass expansion of the first is shown in Fig.~1 where
it is plotted as a function of the dimensionless ratio $k=m/m_c$,
for two different values of the ratio $\gamma=Q^2/M_D^2=2,4$. One
can easily see, that even at relatively large masses of order of
the $b-$quark mass ($k\sim 3$) the heavy mass limit and even the
first$-$order expansion in inverse mass are not reliable.
Including the second $O(1/m^2)$ term one improves the situation
for the $b-$quark, but the approximate result for the moments
still remains very far from  the exact answer in the case of
$c-$quark. We conclude that the heavy quark limit is not a
reliable approximation  for the parton distributions and
fragmentation functions of c-quark. The analysis for higher
moments (second and third) totally confirm this conclusion.

The author thanks  B.L. Ioffe and A.~Khodjamirian for useful
discussions. This work was supported in part  by the Russian
Foundation of Basic Research, project no. 06-02-16905a and the
funds from EC to the project "Study of the Strong Interacting
Matter" under contract N0. R113-CT-2004-506078.

\end{document}